\documentclass{article}
\usepackage{graphicx} 
\usepackage{url}
\usepackage{todonotes}
\usepackage{amsmath}
\newcommand{\pns}{\mathit{pns}}
\newcommand{\pnc}{\mathit{pnc}}
\newcommand{\nonce}{\mathit{nonce}}

\usepackage{authblk}

\usepackage[pageanchor=false,colorlinks,linkcolor=blue,urlcolor=blue,citecolor=blue]{hyperref}

\title{A Brief Note on Cryptographic Pseudonyms for Anonymous Credentials
}
\author[1]{René Mayrhofer\thanks{Full disclosure: These authors are also affiliated with Google, but this paper is published primarily with the academic point of view.}}
\author[2]{Anja Lehmann}
\author[3]{abhi shelat\textsuperscript{*}}
\affil[1]{Johannes Kepler University Linz, Austria}
\affil[2]{Hasso Plattner Institute, University of Potsdam, Germany}
\affil[3]{Northeastern University, USA}
\date{Initial sketch: October 2024, Updated: \today}

\begin{document}

\maketitle

\begin{abstract}
    This paper describes pseudonyms for the upcoming European Identity Wallet (EUDIW) architecture from both a cryptographic and an implementation perspective. Its main goal is to provide technical insights into the achievable properties and cryptographic realizations. In particular, we (1) outline the \emph{security and privacy requirements} of EUDI pseudonyms as the basis for building consensus on the cross-country decision maker level; (2) sketch an \emph{abstract cryptographic protocol} that fulfills these requirements; and (3) suggest \emph{two instantiation options} for the protocol sketch based on well-studied building blocks.
    A complete specification of the formal properties, as well as the specific set of credential issuance, provisioning, and pseudonym presentation generation is outside the scope of this paper, but is expected to follow as future work.
\end{abstract}

\section{Introduction and Definition}
A pseudonym in the EUDIW can be created by the holder of the respective ID (the end user) in the UI of their wallet app. The pseudonym is specific for the relying party (RP) and it is a goal that users can hold multiple different pseudonyms per RP, and that they can select which pseudonym to use for every interaction with that RP (e.g., creating different Amazon accounts to keep family purchases separate from business orders).

The intended use case for pseudonyms as described in this document is using EUDI for private, third-party services such as web sites. These scenarios may benefit from the guarantee that the pseudonym is derived from a real EUDI but without the association to real name or other unique identifiers. From a usability point of view, this may or may not be equivalent to the login process itself; it is conceivable that the EUDI derived pseudonyms are used during creation (and recovery) of a third party (web) service account, while other established mechanisms like Passkeys could be used for the standard login after account creation/activation. These pseudonyms should be easily recoverable by the holder in case of a wallet re-installation, e.g., after loss of the smartphone.

A pseudonym can only be used in scenarios that do not require KYC (`know your customer') or other real name verification by EU regulation or national law.

\section{Motivation and Context}
The difference between such an EUDI pseudonym and an arbitrary \emph{account creation} with fabricated attributes at that RP is that the RP receives a guarantee that the pseudonym is backed by a credential that has been issued to an end user. That is, the purpose is to add support for anonymous, but accountable pseudonyms that can be derived from an existing EUDIW credential. 
The RP must not be able to link the end user's ID to the pseudonym, but they can trust that an end user ID exists behind it. Without that guarantee, we believe there would be no relevant use case for EUDI pseudonyms, because they would be indistinguishable from arbitrary made-up accounts, and the RP could continue to accept such arbitrary username/password based logins and would not have to go through authentication with EUDIW (and the registration/integration/etc. that it entails, which is a cost factor).

It is noteworthy that \emph{account recovery} after the primary login mechanism such as the user password or Passkeys has become invalid or unavailable (e.g., because the user has forgotten their login credentials or their primary device has been lost) is a well-known hard problem for all service providers. Supporting account recovery processes through EUDI wallets would provide a real benefit to the whole web ecosystem, both for users and service providers: users do not have to wade through provider-specific workflows with often unresponsive user support and/or inflexible processes that often result in accounts with important data becoming effectively unrecoverable for them; and service providers do not have to subject themselves to additional attack surface through such recovery procedures.

A secondary goal could be to prevent troll bots from creating a large number of accounts controlled by a small number of people. That is, to provide \emph{rate limited pseudonyms}. There are legitimate use cases (e.g., social media, web forums, etc.) for restricting a single person to a single or a few\footnote{For whatever definition of `few' makes sense to that platform.} accounts on each platform. Using EUDI pseudonyms, if linked to the respective platform, could be an effective technical way to combat the sock puppet/troll account problem (which is a real one in today's democracies).

Pseudonyms only add privacy when coupled with \emph{selective disclosure}. If the holder transmits their real name, a unique ID (such as a citizen or social security number), or a biometric attribute (such as a face picture) as attested by the issuer, they are not pseudonymous. Therefore, using a pseudonym means creating a made-up identifier (e.g., name or username) at the RP and then consistently authenticating with the same identifier.\footnote{Note: If the holder/wallet is not using the consistent identifier with the RP for subsequent interactions, then they are not pseudonymous, but anonymous interactions just relying on the unlinkability property and not transmitting \emph{any} identifying attributes of the main, real ID without the need for creating a pseudonym.} That identifier is local to the RP---either stored at the RP server-side or in the EUDIW---and presented to the RP in every interaction but verified by the RP to belong to a consistent account.

We refer to~\cite{abc4trust} for an overview of such pseudonyms in the context of anonymous credentials. Therein, pseudonyms are categorized as \emph{verifiable} (verifiably derived from a user controlled key), \emph{certified} (verifiably derived from a user holding a certain credential), and \emph{scope-exclusive} (verifiably derived for a certain scope, and enforcing pseudonyms to be unique per scope). In the rest of the document, we consider certified and (mostly) scope-exclusive pseudonyms.

\section{Terminology}
We use the following terminology, explicitly with synonyms (given in brackets) from different literature or application domains.
\begin{itemize}
    \item \textbf{Holder} (\emph{prover, user}): The individual user, who is in control of (selectively) sharing their credential attributes. In the EUDIW architecture, users are not assumed to hold (device independent) user secret keys ($usk$) themselves.
    
    \item \textbf{Wallet}: Technical implementation, respectively, representation of a holder. A wallet is assumed to be executed on a device under physical control by the holder (e.g., a smartphone) and to include a trusted component (Wallet Secure Cryptography Application (WSCA) often implemented as a separate Wallet Secure Cryptography Device (WCSD) such as an embedded secure element) for handling private key material.
    
    \item \textbf{Holder key} (\emph{device secret key} $dsk$): A private key unique to a single wallet respectively WSCD that cannot be extracted or copied. Together with (e.g., biometric) user authentication to ensure user-to-device holder binding, the $dsk$ replaces a user key for most cryptographic uses --- but with the $dsk$ explicitly being non-transferrable between wallets. We will \emph{not} use $dsk$ directly to derive pseudonyms, as this would violate the transferability we require for pseudonyms.
    
    \item \textbf{PID provider} (\emph{issuer}, \emph{issuing authority}): An authority trusted by other parties to establish the link between real user identity and associated attributes.
    
    \item \textbf{Relying Party (RP)} (\emph{verifier}): The (open set of) service providers receiving and verifying/relying on attributes about individual users.
    
    \item \textbf{Credential} (\emph{attestation}): The set of attributes (typically in rich data formats, but often represented as canonicalized strings) signed by the PID provider and (selectively) presented to the RP.
    
    \item \textbf{Scope} ($scp$) (\emph{context} ($ctx$), \emph{basename} ($bsn$), \emph{RP ID $rid$}, also referred to as \emph{directed identity}): A unique identifier for an RP, typically in the form of an integer or string (cf.\ e.g.~\cite{cryptoeprint:2014/708} for more cryptographic details on the scope). Examples of practical use of such scopes already exist, e.g., in the ID-Austria system in the form of \emph{bPK} (`bereichsspezifisches Personenkennzeichen'). In the more general case of web service use, we recommend the $scp$ to be set to the FQDN or URL of the RP.
    \item\textbf{Index} ($idx$) (\emph{tag, counter}). If an upper limit of pseudonyms per scope is desired, this can be steered by an additional index value $idx$. The index can be a simple integer to upper-bound the number of pseudonyms per RP (e.g., for rate-limiting purposes), or be a more complex descriptive string that is meaningful to distinguish these pseudonyms from the holder and/or RP point of view.
\end{itemize}

\section{Security and Privacy Requirements}
The top-level anonymity property requires that pseudonyms must be \emph{unlinkable}  across scopes and indexes, and not leak any information about the identity of the user. This can be defined roughly through this game: 
a relying party (RP) is allowed to query for pseudonyms for any user, $scp$, and $idx$.
At some point, the RP picks two users, Alice and Bob, a given scope $scp$ and index $idx$.  A fair coin is flipped to determine who should register their pseudonym under $(scp,idx)$ and the RP receives the resulting challenge pseudonym.  The RP is (before and after the challenge) allowed to request pseudonyms from Alice and Bob for all scopes and indexes other than the one used for the challenge pseudonym. Eventually, the RP must guess whether it was Alice or Bob who registered the challenge pseudonym, and the RP should succeed with probability close to $\frac{1}{2}$. 
In short, the pseudonym registration process should not leak any information about the identity of the user, not even whether two pseudonyms (for different scope or index) belong to the same user or not. Moreover, this should hold even if the RP and issuer collude.

At the same time, the accountability (\emph{soundness}) property guarantees that a malicious user can only produce certified pseudonyms when they own a corresponding credential. For scope-exclusive pseudonyms, it is further guaranteed that a malicious user can only produce 
1 (or another well-defined number, depending on the use case) pseudonym per scope.
Anonymity and accountability are naturally at odds, so it is interesting that we can achieve both properties at once.

More specifically, we define the following granular requirements. A formal treatment of such pseudonyms can be found, e.g., in~\cite{cryptoeprint:2014/708}.

\begin{enumerate}
    \item \textbf{Soundness}: A holder \textsc{shall} be able to create  
    pseudonyms only when they possess an underlying  EUDI credential. When scope-exclusive pseudonyms (see more details below) are used, soundness further ensures that the holder can only create a unique   
    (or multiple, but potentially limited to a `small' number) pseudonym(s) for each RP they want to interact with, 
    The PID provider must not be able to block a holder from creating a new pseudonym.
    
    Note that this requirement explicitly does not imply that users are free to choose completely arbitrary random key material or identifiers as their pseudonyms, as this would break unforgeability by allowing two users to choose the same key and act in each other's (pseudonymous) name.
    
    \item \textbf{Unforgeability} (\emph{unspoofability, non-frameability}): A malicious holder \textsc{shall not} be able to create a fake ID that passes for a real ID and/or to pass another user's real ID as theirs. This includes both `real name' and pseudonymous IDs: it must be impossible to create a pseudonym (in the sense of this paper) that is not actually linked to a real PID, and it must be impossible to create a pseudonym that is indistinguishable from a different user's real PID. 
    Violating this property would lead to identity theft. 
    
    A stronger property, denoted as \emph{non-frameability}, further ensures that even a malicious \emph{issuer} cannot (re-)create pseudonyms of honest users, i.e., malicious issuers cannot impersonate honest users through pseudonymous IDs.

    One immediate implication of the unforgeability property is that pseudonyms must be distinguishable from real PIDs from an RP point of view to prevent spoofing other holder's real names (or other attributes).
    
    \item \textbf{Unlinkability}: The RP \textsc{shall not} be able to link the pseudonym to the real, unique ID or to other pseudonyms of the same user.

    The RP and PID provider \textsc{shall not} be able to link interactions of the holder with the same or other RPs even under the assumption of collusion between PID provider and RP(s) (also referred to as \emph{untraceability}).
    
    \item \textbf{Unobservability}: The PID provider \textsc{should not} be able to trace/observe the usage of pseudonyms at RPs, and (ideally) \textsc{should not} even be able to determine how many pseudonyms an ID holder is using for each RP.
    
    \item \textbf{Selective disclosure}: The holder \textsc{shall} be able to choose which of the original attributes they share with the RP and which they do not share. 
    An RP can always distinguish disclosed, attested attributes that chain back to the real ID from made-up attributes that are substituted for undisclosed real attributes, because otherwise unforgeability/unspoofability would be violated, and therefore identity theft would be possible.
    
    \item \textbf{Scope-exclusive pseudonyms} (\emph{sybil-resistance}): Some use cases might require an upper bound, or even uniqueness of pseudonyms per RP.
    This has been denoted scope-exclusive pseudonyms, where the RP can---in addition to the soundness property above---be guaranteed that a user can only derive a \emph{unique} pseudonym for a given scope. Pseudonyms across scopes remain unlinkable. 
    As unique pseudonyms  per RP can be too restrictive, we derive the pseudonym over two values: a scope $scp$ and  index $idx$. The scope is typically an RP identifier (e.g., the fully qualified domain name or a unique RP registry number), whereas the index $idx$ is a counter to determine the upper limit of pseudonyms allowed per RP.
    
       In such use cases, a malicious holder \textsc{shall not} be able to create more pseudonyms  than upper-bounded by $idx$ per $scp$.

    \item \textbf{Transferability}: Pseudonyms derived from an EUDI credential \textsc{shall} be transferable from one wallet instance to another (e.g., when the user switches to a new device) and recoverable when the previous wallet instance becomes inoperable. 
\end{enumerate}

Note: KYC use cases are out of scope, because they will not be able to complete without the holder selectively disclosing their full, real name and potentially other unique identifiers of their real PID, which invalidates unlinkability and can therefore not be performed in a pseudonymous manner.

\section{Technical Solution Sketch}
We start by describing the general idea and then sketch different instantiations.

\subsection{High-level Idea}
\paragraph{Step 1: Adding a pseudonym seed to the credential.}
The issued EUDI credential for a holder Alice must contain a new attribute, `pseudonym-seed', or $\pns$ for short, which is a high-entropy and unique value specific to every holder.

\emph{Note}: The EUDIW \textsc{shall} ensure that $\pns$ is never disclosed directly.\footnote{For transferability, the wallet may provide export capability in a secure manner. However, $\pns$ must never be revealed to an adversary.}  When scope-exclusive pseudonyms and/or non-frameability is desired, this must also hold for issuance. 
This effectively implies a setup process by which a cryptographic commitment to $\pns$ is included in the credential instead of the $\pns$ directly, and that only the user holds the opening information for the commitment.

Further, for soundness, it is important that each user can only submit 1 $pns$ to the identity issuer. This can either be realized through maintaining state by the issuer to ensure the same (blinded / commitment over a) $pns$ is used across multiple issuance flows, or relying on blind carried-over attributes~\cite{abc4trust} from existing credentials of the holder.

\paragraph{Step 2: Adding a pseudonym to the presentation.}
For a pseudonymous presentation, this pseudonym-seed is used as a key for a pseudo-random function (PRF). More precisely, the wallet computes a pseudonym as the output of the PRF applied to an RP identifier $scp$ (e.g., the fully qualified domain name or a unique RP registry number) together with an index $idx$ (e.g., a simple integer to upper-bound the number of pseudonyms per RP) to create an RP-specific pseudonym $nym$:
    \[ nym = \mathsf{PRF}_{\pns}(scp||idx) \]

In addition, the holder shares only non-identifying attributes (e.g., their age, or not even that) and the RP-specific identifier $nym$ with the RP and provides a zero knowledge presentation that proves that:
    \begin{enumerate}
        \item The holder has a valid credential and knows the device holder key (when device binding is required).
        \item That credential includes (a commitment to) $\pns$ as one attribute.
        \item The pseudonym ID $nym$ has been created by computing a PRF keyed on $\pns$ with the RP identifier $scp$ (and an optional, possibly hidden, $idx$).
    \end{enumerate}

\emph{Note}: Since $\pns$ (as well as the commitment to the $\pns$) is a unique identifier of the holder, it must never be revealed during presentation. If it is leaked, unlinkability may be broken, even retroactively. This holds in particular for scope-exclusive pseudonyms as sketched above, which are usually deterministically derived values from $\pns, scp$ and $idx$. If only certified (but not scope-exclusive) pseudonyms are used, the (accidental or enforced) exposure of $\pns$ is less critical, as these pseudonyms can also be derived in a perfectly hiding way.  Additionally, when there are multiple RPs, care must be taken in the implementation to ensure that the $scp$ is domain-restricted to an RP.  In other words, one RP cannot request an $scp$ produced by another RP. In practice, this can be done by incorporating information from the TLS session between the wallet and the RP into the $scp$.\footnote{An obvious candidate for verifying $scp$ is the common name attribute in the server (RP) TLS certificate.}

\subsection{Instantiations}
There are different ways to instantiate the PRF and the aforementioned zero-knowledge proof. The following briefly sketches some of the possible solutions.

\subsubsection{Generic ZKP and HMAC}
The aforementioned idea can be realized through a generic circuit-based ZKP, that can be performed on EUDI credentials in form of an \texttt{mdoc} or \texttt{SD-JWT} credential (or any other credential format), possibly signed with classic signature schemes such as ECDSA.

To realize the blind signing of the pseudonym seed $\pns$, the holder computes a pseudonym commitment\footnote{A commitment constructed from SHA-256 typically includes a nonce. For simplicity of notation, this nonce could be incorporated into the $\pns$ string. The length of $\pns$ depends on whether computational or statistical unlinkability is desired. However, for modularity of proof arguments between the commitment to and use of $pns$, this proposal explicitly includes the nonce.}:

     \[\pnc = \mathsf{SHA}\text{-}256(\pns, \nonce) \]

Alternatively, a standard Pedersen commitment~\cite{crypto-1991-1671} with two separate generators $g$ and $h$ (with no particular requirements for the base group $G$ as part of this first instantiation) can be used:

    \[ \pnc=g^{\pns} \cdot h^{\nonce} \] 

The issuer attested/signed real ID credential (the \texttt{mdoc}, \texttt{SD-JWT}, or other representation) includes $\pnc$ in its attributes (if multiple \texttt{mdoc}s are issued, then the primary \texttt{mdoc} used to produce ZK presentations). We recommend every issued credential to include a fresh commitment with a new random $\nonce$ This $\nonce$ will need to be stored within the wallet instance as long as the credential is valid, but does not need to be protected with the same level of confidentiality as the long-term stable $\pns$ itself.

\emph{Note}: Under the assumptions that $\pns$ is never revealed to any other party that is not the holder and their wallet(s), and that the issuer will only accept a single (commitment to) a unique $\pns$ value per holder, we not require the commitment to be (perfectly or just computationally) hiding: the issuer must be able to distinguish if a $\pnc$ during a new enrollment commits to the same $\pns$ the holder has used in potential previous instances, and no other parties get access to either $\pns$ or $\pnc$. The $\nonce$ is therefore not strictly required for the purposes of our protocol sketch and may be omitted in both the hash-based and Pedersen commitment variants for optimization purposes.

\medskip

For instantiating the PRF of the pseudonym function, we recommend a standard HMAC-SHA256 for practical reasons (under the assumption of practical collision resistance of SHA-256)~\cite{10.1007/978-3-662-44371-2_7}:

    \[ nym = \mathsf{HMAC}_{\pns}(scp||idx) \]

When applied on top of an existing \texttt{mdoc} as a particular example, the three parts of the above sketched ZKP proof translate to the statement: ``There exists an MSO that is signed by the issuer that is valid and accompanied by its device-bound transcript signature, which contains an attribute $\pnc$, and $\pnc=\mathsf{SHA}\text{-}256(\pns, \nonce)$\footnote{This could use either the hash-based or the Pedersen commitment variants, depending on which one has been used for the credential provisioning.} for a $\nonce$ known to the holder, and $nym = \mathsf{HMAC}_{\pns}(scp||idx)$''.

Such a statement can be proven under zero knowledge using different signature and/or presentation methods. Details of the issuer signature and presentation format are outside the scope of this paper.
This construction and verification of pseudonyms can, e.g., build upon mdoc-via-ZKP presentation~\cite{cryptoeprint:2024/2010} and only assumes availability of a (certified) ZKP circuit that can include statements about hashes (e.g. SHA-256).

\subsubsection{BBS and Pseudonyms}
BBS is an optimized signature scheme that enables efficient zero-knowledge proofs and also supports pseudonyms.

To include the pseudonym seed $\pns$, one can use the blind issuance feature natively supported by BBS signatures (we simply sign $\pns$ in the form of a `commitment' 

    \[ \pnc=g^{\pns} \] 

where $g$ is a generator for the BBS credential). The BBS credential then contains $\pns$ as a direct attribute, without revealing that value to the issuer. Similarly as above, we can also use full-fledged blind signing by using a fresh commitment $\pnc=g^{\pns} \cdot h^{\nonce}$ for a fresh nonce in every issuance. This has no impact on the derived pseudonyms and presentations, but provides better privacy in issuance.

The PRF for the pseudonym can be instantiated, e.g., through Hash-DH as $\mathsf{PRF}_k(x)=\mathsf{H}(x)^k$ for a hash function $\mathsf{H}$. That is, the holder computes the pseudonym as 

    \[ nym=\mathsf{H}(scp||idx)^{\pns} \] 

and proves that $\pns$ is the same value as in the BBS credential. The latter is a simple Schnorr proof of equality of two discrete logarithms. This pseudonym computation has been ISO-standardized in the context of DAA~\cite{ISODAA}, and currently also undergoes standardization in the IETF for BBS credentials~\cite{BBSnym}, and for Privacy-Pass extensions based on anonymous credentials~\cite{IETFArc,IETFgoogle}. 

The drawback of this simple pseudonym construction is that its unlinkability property relies on the discrete logarithm assumption. That is, adding such pseudonyms reduces perfect privacy of BBS-based ZKPs to computational privacy, which can be retroactively broken when efficient quantum computers become available. Thus, the simple construction should only be used when everlasting privacy is \emph{not} required. Pseudonym constructions with everlasting privacy exist as well, but incur higher---yet still practical---costs~\cite{everlastingnyms}. Finally, we note that the challenge of everlasting privacy occurs only for scope-exclusive pseudonyms. For certified pseudonyms that are not limited by a scope or index, pseudonyms can simply be derived through perfectly hiding (Pedersen) commitments with a corresponding Schnorr-based ZKP.

\subsubsection{Linkable fallback option to integrate with batch issuance (deprecated)} 

Note that this basic concept of including a commitment to $pns$ in the issuer signed credential---particularly the Pedersen variant---could also be applied to older-style batch-issued single-presentation~\cite{bib:2021:iso:18013-5} credentials with a Schnorr-style proof of knowledge of a Pedersen commitment to $\pns$~\cite{cryptoeprint:2024/1444}. This combination can provide soundness, unforgeability, selective disclosure, and scope-exclusive pseudonyms, but only partial unlinkability under very specific assumptions.

However, we explicitly note that full unlinkability (and unobservability) under the assumption of collusion with the respective issuer \textbf{cannot} be achieved with batch-issued credentials without a full ZKP presentation layer such as one of the two previous options. It therefore \textbf{fails the mandatory requirements} defined above.

This inability to achieve full unlinkability is independent of the additional practical disadvantages of batch issuance:
\begin{itemize}
    \item Issuers incur additional cost with batch issuance at scale, because hardware secure module (HSM) signing operations are comparatively costly, and are multiplied by the batch size.
    \item Wallet providers need to implement more complex issuance/provisioning protocols, and need to handle (regular and irregular) re-issuing \emph{before} holders run out of single-use cached credentials.
    \item Holders might be unable to use their credentials if their device is offline when running out of batch credentials. In certain scenarios, this might be a significant concern.
\end{itemize}

All of these are practically relevant problems that might cause the privacy properties of such a solution to fail even more drastically. Concretely, if (batch or single issued) credentials are ever re-used by any of the involved parties, user transactions with these credentials immediately become linkable even within the same RP without assuming any collusion with or breach of the issuer. We consequently strongly recommend against this variant.

\subsection{Practical Transferability of Pseudonyms}
The core of allowing a set of pseudonyms to be transferred from one EUDIW instance to another is transferring the $pns$ after re-provisioning the holder's core EUDI credentials on the new devices.\footnote{If a $\nonce$ has been used for the commitment to $\pnc$, this nonce does not need to be transferred, as we recommend a fresh one to be used for the re-provisioning of the new credential.} As the device-specific key $dsk$ is intentionally not used directly in the derivation of $nym$ (but only for device binding during the presentation), the full set of pseudonyms can be reconstructed given $pns$.

Unfortunately, this creates a clear conflict of interest:
\begin{itemize}
    \item One the one hand, holders should have an easy-to-use mechanism to transfer or reconstruct their EUDI credentials including all derived pseudonyms on a new device.
    \item On other hand, the transfer/recovery mechanism needs to be secure against unintentional misuse, including the threat of holders being coerced (by potentially powerful adversaries) into exporting their $pns$.
\end{itemize}

That is, simply displaying or exporting $pns$ in plaintext and letting the end-user deal with handling it securely is unsafe concerning the latter aspect. The practical handling of transferability is further complicated by the fact that it needs to cover two different scenarios:

\begin{itemize}
    \item Transfer from one EUDIW instance to another when both are online at the same time. This is typical of transferring from an old to a new device with an overlapping transition period. 

    An online transfer is easier to secure, as both devices can participate in e.g. an ephemeral session keyed exchange that never exposes $pns$ in plaintext in transit. UX flows could, e.g., include QRcode or NFC engagement between source and target of such a transfer as specified by mDL engagement~\cite{bib:2021:iso:18013-5} or through explicit confirmation of local WiFi transfers as implemented in the Signal messenger.

    \item Restoring a backup from a previous instance, when the source of the transfer is no longer available at the time of provisioning the target.
\end{itemize}

Solving both scenarios while addressing both aspects described above is difficult in practice. We therefore propose an implementation in two parts:

\begin{enumerate}
    \item For backup and restore at different times and without assuming online availability of the source EUDIW instance, we recommend to rely on existing approaches for E2EE backups such as Android Backup~\cite{Mayrhofer_2021}. These existing services already implement resistance against unintentional leaking of backup data or intentional (including insider and brute-force) attacks on encrypted backups.

    The limitation is that such existing backup solutions generally only allow restoring to a new device within the same platform, e.g., from a previous Android or iOS phone to a new one, but not across platforms.
    
    \item For transferring between different platforms, encrypting it for a specific target as specified in the current draft protocol for Passkeys transfers~\cite{FIDOTransfer} seems the safest approach. We recommend to use this particular protocol when the specification is finished, as this can re-use existing work and implementations are assumed to become widely available.

    The limitation is that a transfer can only be performed as long as the source is still available. Transferring to an unknown future target is not supported.
\end{enumerate}

While this two-part implementation does not cover all potential scenarios, we assume that it will be sufficient for the majority of use cases.

\subsection{Limiting the Number of Pseudonyms}
The number of pseudonyms a holder can create per RP is limited through the index $idx$. Setting $idx$ to a well-defined `empty' value enforces a unique pseudonym per RP. Enabling any value for $idx$ allows the holder to create arbitrary many unlinkable pseudonyms per RP. The RP can also define a certain upper bound $\ell$ and allow the user to have to up $\ell$ unlinkable pseudonyms. This has recently been proposed for anonymous rate limiting, where users should be able to anonymously use a service up to an RP-specified upper bound $\ell$~\cite{IETFArc,IETFgoogle}.  
 
 \paragraph{Generic ZKP.} 
 When the RP receives a pseudonym  $nym = \mathsf{PRF}_{\pns}(scp||idx)$ for $scp||idx$, it must additionally check that $1\leq idx \leq \ell$. 
 The additional overhead to verify one comparison operation is easy to add to the arithmetic circuit.

 \paragraph{BBS.} Supporting a rate limit for BBS requires the same additional step.
 As above, instead of revealing $idx$ and proving well-formedness of $nym = \mathsf{PRF}_{\pns}(scp||idx)$ for $scp||idx$, the holder only reveals $scp$  and additionally proves that the pseudonym is correct for some $idx$ which satisfies $1\leq idx \leq \ell$. To enable such proofs, the Dodis-Yampolskiy PRF is used instead of Hash-DH, and the ZKP of the holder must also include a range proof for $idx$. 
This has first been proposed in~\cite{CloneWars}, and has  recently been adopted for the use in rate-limited anonymous credentials~\cite{IETFgoogle,GoogleBBS}.


\subsection{Further Notes and Open Questions}
\begin{enumerate}
    \item $nym$ can change periodically for a specific site if it wants to eliminate correlating identities over time. This can be done by setting $idx$ to a certain time interval, e.g., day or week. The user then has a persistent pseudonym during that time at the RP, but is unlinkable across time epochs.

    \item The main practical difficulties we see with the solution sketched in this paper are:
    \begin{enumerate}
        \item Managing the secret value $\pns$ on the holder’s device: The anonymity properties rely crucially on the secrecy of $\pns$. However, we also need to sync $\pns$ across Alice’s devices. 

        See above for a discussion on backup/restore or transfer of $pns$.
        
        

        \item The identity issuer needs to ensure that every holder only has a single $\pns$. This can either be done through blindly carrying over this attribute from a seed credential, or letting the issuer keep track of $\pns$ (or rather its committed form $\pnc$). If the issuer already keeps state for rate-limiting purposes, then possibly $\pnc$ can be added to the existing implementation.

        The corollary is that, if a user has $k$ different $\pns$ values, they can therefore create $k \cdot idx$ different pseudonyms for the same site $scp$ instead of being limited to the $idx$ handling.

    \end{enumerate}
\end{enumerate}

\section*{Acknowledgments}
We would like to explicitly thank Peter Lee Altmann, Christian Bormann, Sietse Ringers, and Dirk Balfanz for discussions and feedback on earlier versions of this paper.

This work has been carried out within the scope of Digidow, the Christian Doppler Laboratory for Private Digital Authentication in the Physical World.
We gratefully acknowledge financial support by the Austrian Federal Ministry of Economy, Energy and Tourism, the National Foundation for Research, Technology and Development, the Christian Doppler Research Association, 3 Banken IT GmbH, ekey biometric systems GmbH, Kepler Universit\"atsklinikum GmbH, NXP Semiconductors Austria GmbH \& Co KG, and \"Osterreichische Staatsdruckerei GmbH.

\bibliographystyle{acm}
\bibliography{literature}

\begin{thebibliography}{10}

\bibitem{abc4trust}
{\sc Bichsel, P., Camenisch, J., Dubovitskaya, M., Enderlein, R.~R., Krenn, S.,
  Krontiris, I., Lehmann, A., Neven, G., Nielsen, J.~D., Paquin, C., Preiss,
  F.-S., Rannenberg, K., Sabouri, A., and Stausholm, M.}
\newblock {D2.2 - Architecture for Attribute-based Credential Technologies}.
\newblock \url{https://abc4trust.eu/download/Deliverable_D2.2.pdf}, 2014.

\bibitem{CloneWars}
{\sc Camenisch, J., Hohenberger, S., Kohlweiss, M., Lysyanskaya, A., and
  Meyerovich, M.}
\newblock {How to Win the Clone Wars: Efficient Periodic n-Times Anonymous
  Authentication}.
\newblock ACM CCS 2006. \url{https://eprint.iacr.org/2006/454}, 2006.

\bibitem{cryptoeprint:2014/708}
{\sc Camenisch, J., Krenn, S., Lehmann, A., Mikkelsen, G.~L., Neven, G., and
  Østergaard Pedersen, M.}
\newblock Formal treatment of privacy-enhancing credential systems.
\newblock Cryptology {ePrint} Archive, Paper 2014/708.
  \url{https://eprint.iacr.org/2014/708}, 2014.

\bibitem{everlastingnyms}
{\sc Chairattana-Apirom, R., Döttling, N., Lysyanskaya, A., and Tessaro, S.}
\newblock {Everlasting Anonymous Rate-Limited Tokens}.
\newblock to appear at Asiacrypt 2025. \url{https://eprint.iacr.org/2025/1030},
  2025.

\bibitem{cryptoeprint:2024/2010}
{\sc Frigo, M., and abhi shelat}.
\newblock Anonymous credentials from {ECDSA}.
\newblock Cryptology {ePrint} Archive, Paper 2024/2010.
  \url{https://eprint.iacr.org/2024/2010}, 2024.

\bibitem{10.1007/978-3-662-44371-2_7}
{\sc Ga{\v{z}}i, P., Pietrzak, K., and Ryb{\'a}r, M.}
\newblock The exact prf-security of nmac and hmac.
\newblock In {\em Advances in Cryptology -- CRYPTO 2014\/} (Berlin, Heidelberg,
  2014), J.~A. Garay and R.~Gennaro, Eds., Springer Berlin Heidelberg,
  pp.~113--130.

\bibitem{ISODAA}
{\sc ISO}.
\newblock Iso 20008-2:2013: Information technology - security techniques -
  anonymous digital signatures - part 2: Mechanisms using a group public key.
\newblock Standard, International Organization for Standardization, Nov. 2013.

\bibitem{bib:2021:iso:18013-5}
{\sc {ISO/IEC 18013-5:2021}}.
\newblock {Personal identification --- ISO-compliant driving licence --- Part
  5: Mobile driving licence (mDL) application}.
\newblock {International Standard}, International Organization for
  Standardization, 2021.

\bibitem{BBSnym}
{\sc Kalos, V., and Bernstein, G.~M.}
\newblock {BBS} per verifier linkability.
\newblock
  \url{https://datatracker.ietf.org/doc/draft-irtf-cfrg-bbs-per-verifier-linkability},
  2025.

\bibitem{GoogleBBS}
{\sc Katz, J., Raykova, M., and Schlesinger, S.}
\newblock Anonymous credentials with range proofs and rate limiting.
\newblock
  \url{https://docs.rs/crate/authenticated-pseudonyms/latest/source/design/Range.pdf},
  2025.

\bibitem{Mayrhofer_2021}
{\sc Mayrhofer, R., Stoep, J.~V., Brubaker, C., and Kralevich, N.}
\newblock The android platform security model.
\newblock {\em ACM Transactions on Privacy and Security 24}, 3 (Apr. 2021),
  1–35.

\bibitem{crypto-1991-1671}
{\sc Pedersen, T.~P.}
\newblock Non-interactive and information-theoretic secure verifiable secret
  sharing.
\newblock In {\em Advances in Cryptology - CRYPTO '91, 11th Annual
  International Cryptology Conference, Santa Barbara, California, USA, August
  11-15, 1991, Proceedings\/} (1991), vol.~576 of {\em Lecture Notes in
  Computer Science}, Springer, pp.~129--140.

\bibitem{IETFgoogle}
{\sc Schlesinger, S., and Katz, J.}
\newblock Anonymous credit tokens.
\newblock
  \url{https://samuelschlesinger.github.io/ietf-anonymous-credit-tokens/draft-schlesinger-cfrg-act.html},
  2025.

\bibitem{FIDOTransfer}
{\sc Steele, N., Islam, R., Åberg, A., Léveillé, R., Hinton, O., Salamon,
  J., Bedair, A., Campbell, L., and Bajwa, R.}
\newblock Credential exchange protocol.
\newblock \url{https://fidoalliance.org/specs/cx/cxp-v1.0-wd-20240522.html/},
  May 2024.

\bibitem{cryptoeprint:2024/1444}
{\sc Verheul, E.}
\newblock Attestation proof of association – provability that attestation
  keys are bound to the same hardware and person.
\newblock Cryptology {ePrint} Archive, Paper 2024/1444, 2024.

\bibitem{IETFArc}
{\sc Wood, C.~A., and Yun, C.}
\newblock Anonymous rate-limited credentials.
\newblock \url{https://datatracker.ietf.org/doc/draft-yun-cfrg-arc/}, 2025.

\end{thebibliography}

\end{document}